\documentclass[sort&compress,final]{aipproc}
\layoutstyle{8x11single}

\usepackage{psfig,amsfonts,amsmath,amssymb,graphicx,natbib,lscape,nicefrac}

\newcommand{\mum}{$\rm{\mu}$m}
\newcommand{\spitzer}{\textit{Spitzer}}
\newcommand\ion[2]{#1$\;${\scshape{#2}}}

\newcommand\arcsec{\mbox{$^{\prime\prime}$}}%
\newcommand{\apj}{ApJ}
\newcommand{\apjs}{ApJS}
\newcommand{\apjl}{ApJL}
\newcommand{\aap}{A{\&}A}

\newcommand{\mnras}{MNRAS}

\newcommand{\araa}{ARAA}

\begin{document}

\title{Exploring the Galaxy Mass-Metallicity Relation at $z\sim 3-5$}

\author{Tanmoy Laskar}{address={Harvard-Smithsonian Center for Astrophysics, 60
Garden Street, Cambridge, MA 02138}}
\author{Edo Berger}{address={Harvard-Smithsonian Center for Astrophysics, 60
Garden Street, Cambridge, MA 02138}}
\author{Ranga-Ram Chary}{address={Spitzer Science Center, California Institute of
Technology, Pasadena, CA 91125}}


\begin{abstract}
Long-duration gamma-ray bursts (GRBs) provide a premier tool for
studying high-redshift star-forming galaxies thanks to their extreme
brightness and association with massive stars.  Here we use GRBs to
study the galaxy mass-metallicity ($M_*$-$Z$) relation at $z\sim 3-5$,
where conventional direct metallicity measurements are extremely
challenging. We use the interstellar medium metallicities of long-duration 
GRB hosts derived from afterglow absorption spectroscopy ($Z\approx
0.01-1$ Z$_\odot$), in conjunction with host galaxy stellar masses
determined from deep \spitzer\ 3.6 $\mu$m observations of 20 GRB
hosts.  We detect about 1/4 of the hosts with $M_{\rm AB}(I)\approx
-21.5$ to $-22.5$ mag, and place a limit of $M_{\rm AB}(I)\gtrsim
-19$ mag on the remaining hosts from a stacking analysis.  
Using a conservative range of mass-to-light ratios for
simple stellar populations (with ages of 70 Myr to $\sim 2$ Gyr), we
infer the host stellar masses and present the galaxy mass-metallicity
measurements at $z\sim 3-5$ ($\langle z\rangle \approx 3.5$).  We find
that the detected GRB hosts, with $M_*\approx 2\times 10^{10}$
M$_\odot$, display a wide range of metallicities, but that the mean
metallicity at this mass scale, $Z\approx 0.1$ Z$_\odot$, is lower
than measurements at $z\lesssim 3$.  Combined with stacking of the
non-detected hosts (with $M_*\lesssim 4\times 10^9$ M$_\odot$ and
$Z\lesssim 0.03$ Z$_\odot$), we find evidence for the existence of an
$M_*$-$Z$ relation at $z\sim 3.5$ and continued evolution of this
relation to systematically lower metallicities from $z\sim 2$.
\end{abstract}

\maketitle

\section{Introduction}

The simple ``closed-box'' model of galaxy evolution \citep{Talbot1971}
predicts a correlation between the stellar mass and the gas-phase
metallicity of a galaxy (the $M_*$-$Z$ relation).
Such a relation is indeed seen
in the nearby universe \citep{Tremonti2004,Kewley2008,Savaglio2005}.
and is found to be in place out to $z\sim2$ \citep{Erb2006},
progressively evolving to lower metallicities.
Tracing the $M_*$-$Z$ relation and its evolution to even earlier times
will provide insight into the earliest epochs of galaxy evolution,
while allowing us to probe the relative importance of various
galactic-scale phenomena proposed at $z\lesssim 3$.  
However, studying the $M_*$-$Z$ relation beyond $z\sim 3.5$ is challenging
because the nebular emission lines required for robust metallicity
measurements (e.g., H$\alpha$, H$\beta$,
\ion{N}{ii} $\lambda 6583$, \ion{O}{iii} $\lambda\lambda 4959,5007$,
\ion{O}{ii} $\lambda\lambda 3726,3729$) shift into different near- and mid-IR wavebands,
where existing spectrographs have reduced sensitivity compared to the
optical band. This is aggravated by the faintness of the galaxies,
making individual abundance measurements nearly impossible.

An alternative way to determine metallicities at $z\gtrsim 3$ (and in
principle at $z\sim 10$ and beyond \cite{Salvaterra2009,Tanvir2009}) is
absorption spectroscopy of gamma-ray burst (GRB) optical/near-IR
afterglows.  Long-duration GRBs are known to be associated with the
deaths of massive stars (e.g. \cite{Woosley2006}), and therefore
with sites of active star formation.  The large optical luminosities
of GRB afterglows and their intrinsic featureless spectra
provide a unique way to measure interstellar medium (ISM)
metallicities for galaxies at $z\gtrsim 2$ from rest-frame ultraviolet
metal absorption lines and Ly$\alpha$ absorption.  Since the afterglows are
significantly brighter than the underlying host galaxies, this
technique allows us to measure metallicities independent of the galaxy
brightness.  This approach has now been exploited
at least to $z\sim 5$ using optical spectra (e.g.,
\cite{Berger2006,Prochaska2007,Fynbo2009}), and with near-IR spectrographs it
can be implemented to $z\sim 20$.

Naturally, to explore the $M_*$-$Z$ relation at $z\gtrsim 3$ we also
require a determination of the GRB host galaxy stellar masses, and
hence follow-up infrared observations with the \spitzer\ Space
Telescope to probe the rest-frame optical luminosity. We
present the first large set of \spitzer\ observations for GRB host
galaxies at $z\sim 3-5$, and combine the inferred masses with measured
metallicities to explore the $M_*$-$Z$ relation beyond $z\sim 3$.
A full analysis is presented in Laskar et al. (2011), to be submitted to ApJ.

\section{GRB Sample and Data Analysis} 
\label{sec:sample}

We obtained deep observations (about 2 hr per target) of 20 long-duration
GRB host galaxies in the range $z\approx
3-5.8$ using the 3.6 \mum\ band of the
Infra-Red Array Camera (IRAC; \cite{Fazio2004}) on-board the {\it
Spitzer} Space Telescope. The effective
wavelength of the IRAC 3.6 \mum\ band probes the rest-frame spectral
energy distribution (SED) redward of about 5500 \AA\
and therefore provides a robust measure of the stellar mass.

We used optical afterglow images to perform relative astrometry on the
\spitzer\ mosaics and to locate the GRB hosts. 
At the depth of our observations, \spitzer\ images are
confusion-limited for faint sources, requiring us to
model and subtract nearby contaminating sources in 3 cases.
We detect 5 GRB hosts at 3.6 \mum\ (GRBs 050319, 050814, 060707, 
060210, and 060926), with flux densities ranging from about 0.55 to 1.65
$\mu$Jy, with a typical upper limit of about 0.25 $\mu$Jy
(3$\sigma$).
To assess the typical flux density of the non-detected hosts we carry
out a stacking analysis of 11 non-detections with accurate relative astrometry. 
The stacked image does not show a
detection down to a 3$\sigma$ limit of $\lesssim 80$ nJy.

\begin{figure}[ht]
\begin{minipage}{0.24\columnwidth}
\includegraphics[width=\columnwidth,angle=-90]{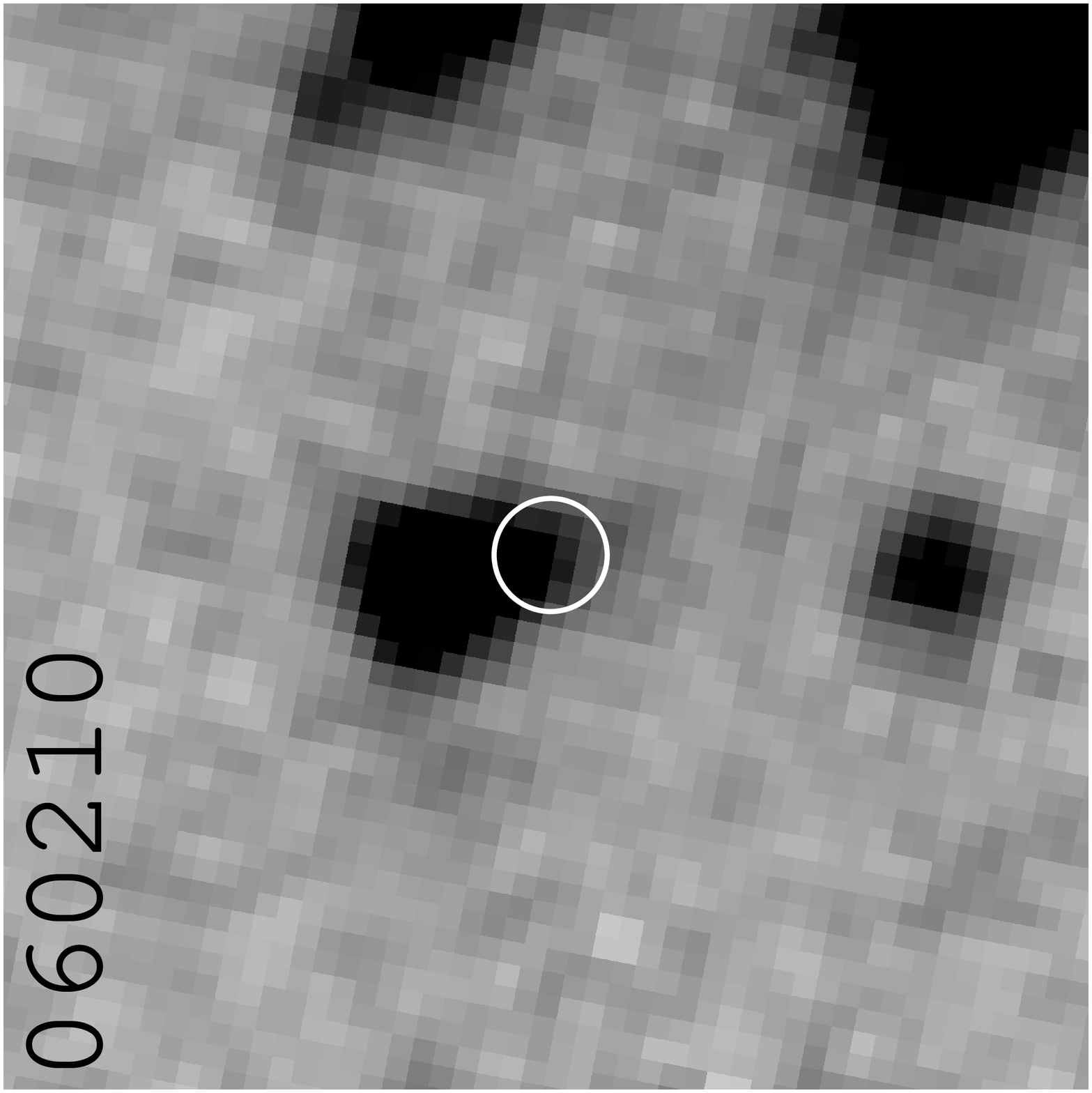}
\end{minipage}
\begin{minipage}{0.24\columnwidth}
\includegraphics[width=\columnwidth,angle=-90]{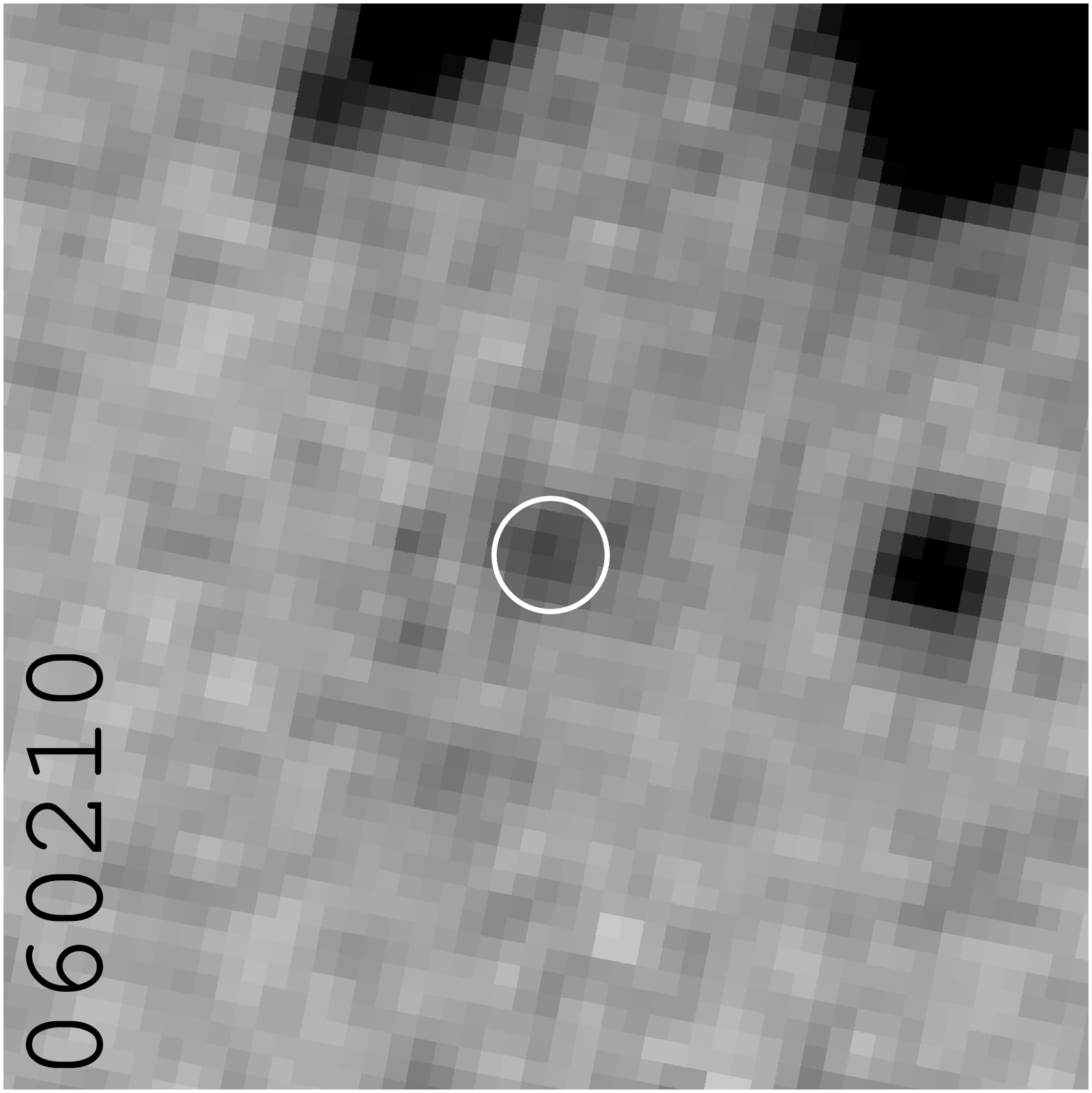}
\end{minipage}\\
\caption{\spitzer\ 3.6 \mum\ detection of the host galaxy of GRB 060210. 
Left: mosaic. Right: contaminating source subtracted.
The circle (1\arcsec\ radius) marks the afterglow position.
North is up and East is to the left, while the pixels are 0\arcsec.4 on a side.  
\label{fig:cutouts1}}
\end{figure}
 
\subsection{Stellar Masses of GRB Hosts at $z\sim 3-5$}
\label{sec:mass}

Computing stellar masses from observed luminosities in a
given wave-band requires knowledge of the mass-to-light ratio and
hence the stellar population age and metallicity.  When multi-band
photometry is available, modeling of the spectral energy distribution
(SED) using stellar population synthesis models can be used to determine
stellar masses, provided that a single stellar population is assumed.
When multi-band photometry is not available, the resulting uncertainty
in the mass-to-light ratio (e.g., at $\sim 1$ \mum) is about an order
of magnitude (e.g., \cite{Magdis2010}).

Here, since we lack broad-band photometry, we determine a range of
mass-to-light ratios for each galaxy in the observed 3.6 \mum\ band
using a wide range of population ages and the single stellar
population models of \citet{Maraston2005} with a Salpeter IMF and
assuming an instantaneous burst of star formation (e-folding time, $\tau=0$).
As expected, the 3.6 \mum\ mass-to-light ratio for these models increases
with stellar population age beyond $\sim 10$ Myr. An upper
bound on the mass-to-light ratio is achieved by setting the stellar
population age to the age of the universe at each host redshift
($\approx 1.8$ Gyr at the median redshift of our sample).  We stress
that this leads to a very conservative maximum mass for each host
galaxy since studies of Lyman Break Galaxies (LBGs) and
Ly$\alpha$ emitters (LAEs) at $z\gtrsim 3$ indicate typical population ages
of $\sim 0.1-0.6$ Gyr \citep{Shapley2005,Reddy2006,Magdis2010,Ono2010}.
The median age for long-duration GRB hosts at $z\sim 1$ of about 70 
Myr \citep{Leibler2010} is probably more typical.
The variation in mass-to-light ratio between
these age values is about an order of magnitude.

The maximal masses inferred
for our sample are $(2.5-5.8)\times 10^{10}$ M$_\odot$, while the
typical (maximal) upper limits are $\lesssim 9\times 10^{9}$
M$_\odot$. The masses inferred for a 70 Myr old population
are about $(0.6-1.4)\times 10^{10}$ M$_\odot$, with typical upper
limits of $\lesssim 2\times 10^9$ M$_\odot$.  The mass limit from
the stack of 11 GRB hosts is $\lesssim 7\times 10^8$ M$_\odot$ for a
70 Myr old population, and $\lesssim 3\times 10^9$ M$_\odot$ for the
maximal age. We also include a literature sample of 5 long-duration GRB hosts
at $z\gtrsim 3$, incorporating two detections with maximal
masses of $1.4\times 10^{10}$ M$_\odot$ (GRB\,060510B;
\cite{Chary2007a}) and $6.7\times 10^{11}$ M$_\odot$ (GRB\,080607;
\cite{Chen2010}).

\section{The Mass-Metallicity Relation at $z\sim 3-5$}
\label{sec:mz}
A typical optical afterglow spectrum 
exhibits a wide range of ISM absorption features
due to rest-frame UV transitions of low- and high-ionization metal
species, which allow a direct determination of the column density of
these elements along the GRB line of sight through the host galaxy.
Combined with a determination of the neutral hydrogen column density
via the Ly$\alpha$ line, it is possible to determine the ISM
abundances (e.g., \cite{Berger2006,Prochaska2007,Fynbo2009}). 
We use \ion{S}{ii}, when available, as a measure of the metallicity, 
primarily since sulfur is not strongly depleted onto dust.
We also place lower limits on the metallicity using
\ion{Si}{ii}, \ion{Si}{iv} and \ion{C}{ii} detections reported by \citet{Fynbo2009}.

Of the 18 GRBs in our sample, six have determined [S/H] values, 
five have no metallicity information, and six have upper limits on their
metallicity from non-detections of \ion{S}{ii} as well as
lower limits based on \ion{Si}{ii} or \ion{Si}{iv} detections.
For GRB\,050908, the metallicity upper limit ($Z<10^2$ Z$_{\odot}$)
is not meaningful and we only report a lower limit based on a \ion{C}{ii} detection.
Using these values we find a wide range of
metallicities (using the Solar abundance values from
\citet{Asplund2005}): $Z\approx 0.01-1.5$ Z$_\odot$ for the {\it
Spitzer}-detected GRB hosts, which have stellar masses of $\sim
2\times 10^{10}$ M$_\odot$. 
We divide the GRBs with available metallicity information
(either a metallicity detection or a bounded range) into two mass bins 
--- the 3.6 \mum\ detections with $M_*\sim2\times 10^{10}$ M$_{\odot}$ 
(Group 1) and the objects included in the stack (Group 2). 
For each group, we perform a Monte Carlo simulation to estimate the mean metallicity, 
yielding $\langle Z_1\rangle =-1.01 \pm 0.17$ and
$\langle Z_2\rangle =-1.52 \pm 0.12$ ($1\sigma$ intervals).
For Group 1, the mean of the maximum inferred stellar masses is $3.7\times 10^{10}$ M$_{\odot}$,
while that of the masses inferred from the 70 Myr populations is $8.8\times 10^9$ M$_{\odot}$.
To obtain mass estimates for Group 2, we scale our stack limit obtained for 
11 non-detections by $\sqrt{11/7}$. Using the mean maximum mass-to-light ratio
of the objects in Group 2 yields an upper limit on the mean stellar mass of
these seven objects of $3.7\times 10^9 M_{\odot}$, 
while using the mean mass-to-light ratio at 70 Myr yields a mass limit of $9.4\times 10^8M_{\odot}$.

In Figure 2, we present the
absorption line metallicities plotted versus the stellar masses
inferred from our \spitzer\ observations and the literature studies.
The 6 GRB hosts that do not have any constraints on their ISM
metallicity are excluded from this plot. 
We find that our two points fall below
the observed relations at $z\lesssim 3.5$, providing evidence that the
galaxy $M_*$-$Z$ relation continues to evolve at $z\sim 3-5$, 
with our stack range probing a somewhat lower mass scale than 
the LBG studies at $z\sim 3.1-3.5$ \citep{Maiolino2008,Mannucci2009}.

\begin{figure}[h]
\centering
\includegraphics[width=0.67\columnwidth]{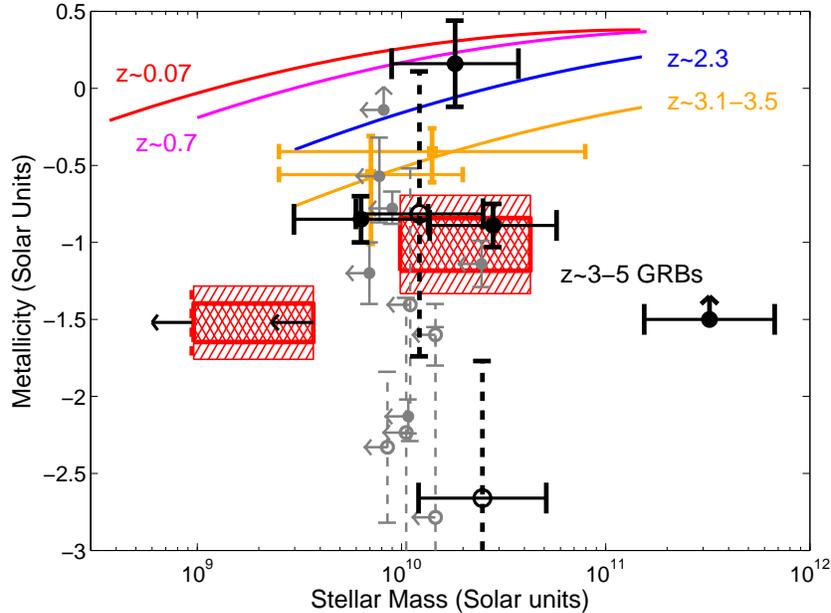}
\caption{Stellar mass plotted as a function of ISM metallicity for our
sample and the 5 previously observed hosts (black=detections;
gray=limits). Confirmed metallicities 
(and one metallicity lower limit) are indicated by filled symbols,
while metallicity ranges are shown by dashed vertical lines 
with open symbols.
The hatched regions designate $1\sigma$ and $2\sigma$ 
intervals for estimates of the mean metallicity at two mass bins of 
$\sim 2\times 10^{10}$ M$_\odot$ (3.6 $\mu$m detections) and
$\lesssim 3.7\times 10^9$ M$_\odot$ (scaled stack limit - see text).
The dashed vertical line indicates the upper limit on the mean
mass of the stack for a 70 Myr population.
These data are consistent with a decline in
metallicity with lower stellar mass --- an $M_*$-$Z$ relation.  Also
shown are the relations for $z\sim 0.07$ \citep{Kewley2008}, $z\sim
0.7$ \citep{Savaglio2005}, $z\sim 2.3$ \citep{Erb2006}, and $z\sim
3.1-3.5$ \citep[][filled squares]{Maiolino2008,Mannucci2009}; the curves at $z\lesssim
2.3$ are $M_*$-$Z$ relations re-calibrated by \citet{Maiolino2008}.  Our two
points at $z\sim 3-5$ fall below these relations, suggesting that the
$M_*$-$Z$ relation continues to evolve to $z\sim 4$.
\label{fig:MZ}}
\end{figure}

\section{Discussion and Conclusions}
\label{sec:conc}

We present the first study of the galaxy mass-metallicity relation at
redshifts of $z\sim 3-5$ using GRB afterglow absorption metallicities
and \spitzer\ follow-up observations.  Five of the 20 GRB hosts in our
sample are detected above a 3$\sigma$ flux density threshold of 0.25
$\mu$Jy, corresponding to a typical stellar mass of $\sim 2\times
10^{10}$ M$_\odot$.  We further place a limit of $\lesssim 3\times
10^9$ M$_\odot$ on the non-detected hosts based on a stacking
analysis.

Using the metallicities of the host galaxies (inferred mainly from
\ion{S}{ii}) we find a range of $\sim 10^2$ at a fixed stellar mass of
$\sim 2\times 10^{10}$ M$_\odot$. The mean metallicity at
this mass scale is about $0.1$ Z$_\odot$.  The mean metallicity
associated with 7 of the 11 the non-detected hosts, which have an upper limit of
$\lesssim 3.7\times 10^9$ M$_\odot$, is $Z\lesssim 0.03$
Z$_\odot$.  Thus, there appears to be an overall decline in
metallicity with decreasing mass, a hint of an $M_*$-$Z$ relation.
Furthermore, our two points on the $M_*$-$Z$ relation lie below the
relations at lower redshifts, suggesting that the relation continues
to evolve at least to $z \sim 4$.  Clearly, additional observations are
required to confirm and increase the statistical significance of this
result.  A sample of 20 additional GRBs at $z\gtrsim 3$ from 2007
through the present is available for study.  This will allow us to
double the existing sample.

The {\it James Webb Space Telescope} (JWST)
will provide a much deeper view of the $M_*$-$Z$ relation at high
redshift.  For instance, the NIRCam instrument on JWST 
will allow us to detect GRB hosts at
$z\sim 3$ down to a mass of $\sim 10^8$ M$_\odot$, and at $z\sim 6$ to
$\sim 3\times 10^8$ M$_\odot$ with similar integration times as our observations.
Equally important, the NIRSpec instrument ($1-5$
\mum) will allow us to determine emission-line metallicities for some
of the hosts, and hence to cross-calibrate the afterglow absorption
metallicities.  With on-going afterglow absorption metallicity
measurements, the GRB sample will continue to play a key role in our
study of high redshift galaxies.

\bibliographystyle{aipproc}

\end{document}